\newcommand*\chem[1]{\ensuremath{\mathrm{#1}}}
\documentclass[12pt, twocolumn]{openjournal}

\usepackage{lipsum}
\usepackage{amssymb}
\usepackage{booktabs}
\usepackage{gensymb}

\usepackage{xcolor}
\usepackage{textgreek}
\usepackage[utf8]{inputenc}
\usepackage[english]{babel}
\usepackage{hyperref}
\hypersetup{
    colorlinks=true,
    linkcolor=blue,
    filecolor=blue,
    urlcolor=red,
    citecolor=blue,
}
\usepackage{color,colortbl}
\usepackage{tensind}
\tensordelimiter{?}
\DeclareGraphicsExtensions{.bmp,.png,.jpg,.pdf}
\usepackage{verbatim}
\usepackage[normalem]{ulem}
\usepackage{orcidlink}
\usepackage{soul}

\graphicspath{ {./figs/} }

\usepackage{amssymb}
\usepackage{amsfonts}
\usepackage{amsmath}
\usepackage{float}

\usepackage[figuresright]{rotating}

\begin{document}


\title{Insights into Lunar Mineralogy: An Unsupervised Approach for Clustering of the Moon Mineral Mapper (M3) spectral data}

\author{Freja Thoresen$^{1,*}$ \orcidlink{0000-0003-2569-550X}}
\author{Igor Drozdovskiy$^{2}$ \orcidlink{0000-0002-5085-1319}}
\author{Aidan Cowley$^{2}$\orcidlink{0000-0001-8692-6207}}
\author{Magdelena Laban$^{2,3}$\orcidlink{0009-0000-0919-757}}
\author{Sebastien Besse$^{4}$\orcidlink{0000-0002-1052-5439}}
\author{Sylvain Blunier$^{2}$\orcidlink{0000-0002-9456-2332}}

\thanks{$^*$E-mail: frejathoresen@gmail.com}
\affiliation{$^{1}$ Directorate of Human and Robotics Exploration, European Space Agency (ESA), European Space Research and Technology Centre (ESTEC), Noordwijk,Netherlands\\
$^{2}$Directorate of Human and Robotics Exploration, European Space Agency (ESA), European Astronaut Centre (EAC), Cologne, Germany\\
$^{3}$University of Wroclaw, Faculty of Earth Sciences and Environmental Management, Institute of Geological Sciences, Wroclaw, Poland\\
$^{4}$Directorate of Science, European Space Agency (ESA), European Space Astronomy Centre (ESAC), Madrid, Spain
}

\begin{abstract}
    This paper presents a novel method for mapping spectral features of the Moon using machine learning-based clustering of hyperspectral data from the Moon Mineral Mapper (M3) imaging spectrometer. The method uses a convolutional variational autoencoder to reduce the dimensionality of the spectral data and extract features of the spectra. Then, a k-means algorithm is applied to cluster the latent variables into five distinct groups, corresponding to dominant spectral features, which are related to the mineral composition of the Moon’s surface. The resulting global spectral cluster map shows the distribution of the five clusters on the Moon, which consist of a mixture of, among others, plagioclase, pyroxene, olivine, and Fe-bearing minerals across the Moon's surface. The clusters are compared to the mineral maps from the Kaguya mission, which showed that the locations of the clusters overlap with the locations of high wt\% of minerals such as plagioclase, clinopyroxene, and olivine. The paper demonstrates the usefulness of unbiased unsupervised learning for lunar mineral exploration and provides a comprehensive analysis of lunar mineralogy.
\end{abstract}

\section{Introduction}
Recent efforts in the space community have been focused on exploring and understanding the mineralogy of the Moon. Mapping the distribution and abundance of various mineralogical resources is valuable, not only for scientific exploration, but also for potential future resource prospecting (as a part of the in-situ resource utilization (ISRU) efforts) \cite{Anand2012}. 

The lunar surface is rich in chemical elements such as iron (pyroxenes, olivines, ilmenite), aluminum (plagioclase feldspar), titanium (ilmenite), oxygen (plagioclase feldspars, pyroxene, olivines, ilmenite), silicon (plagioclase feldspars, pyroxenes, olivines), and calcium (plagioclase feldspars, pyroxenes) \cite{Heiken1991}. Providing an easy-access method for mineral mapping and recognition is therefore a crucial part of Moon exploration. 

Hyperspectral imaging (HSI) is sensitive to chemical and physical properties of a planetary surface and delivers a wide range of multi-dimensional information about the observed materials encoded in reflected spectral features. HSI can therefore be used to estimate mineral compositions on the surface of the Moon by spectral absorption features. The Moon Mineral Mapper (M3) is a spectrometer from the Chandrayaan-1  mission, which provides HSI in a wide range of wavelengths in visible to near-infrared light \cite{Pieters2009}. 

Traditionally, an empirical equation is used when determining if a mineral is present in a spectral dataset \cite{Pieters2002}. However, in recent years machine learning (ML) methods have shown to be useful for mineral classifications \cite{Jahoda2021}. Unlike supervised classification, which predetermines the labeling of the data, clustering does not explicitly define the groups or labels in advance. The unsupervised learning approaches are employed to create a feature space in which the data can be effectively clustered \cite{Xie2015}, potentially revealing hidden structural patterns in measurements that could be missed with supervised classification. However, among the challenges of this approach is the interpretability of the revealed patterns, which are addressed in a follow-up investigation combined with analytical techniques. 

In this study, we present a mapping and a comprehensive analysis of the spectral features of the Moon using the M3 dataset. Using machine learning-based clustering methods, we identified the dominant spatial patterns corresponding to distributions of mixed mineralogies on a global scale of the Moon's surface. We then introduce a method to identify the spectroscopic dominant mineral groups from the perspective of spectral visible to near-infrared measurements. 
The analysis does not rely on any assumptions on the mineralogy of the Moon. Hence, the number of clusters, as well as the clustering approach are based on an unsupervised identification of the ML features. 

\subsection{Mineralogy of the Moon}

The Moon’s surface geology can be divided into two main lithological types: basaltic maria covering ~17 \% of the Moon’s surface and feldspathic highlands dominating the rest of the lunar surface \cite{Head1976}. As a result of the high rate of impacts, the two main types of igneous rocks are heavily mixed with impactites at various degrees due to the different exposure times to the space weathering \cite{Lucey2006}. 

The predominantly old primary lunar crust is mostly of anorthositic composition, as well as noritic, gabbroic, and troctolitic \cite{Tompkins1999}. The mature highland rocks, predominantly anorthosites, are often of noritic composition, with scarce amounts of orthopyroxenes and olivines \cite{Pieters1986}. The lunar lowland “mare” regions, on the other hand, are composed of basaltic rocks that include pyroxenes, olivines, metal oxides such as  (Fe-, Mg-, Al-, Ti- \& Cr-) spinels, rutile or ilmenite with addition of quartz and cristobalite \cite{Che2021}. Globally, major lunar rock-forming minerals are plagioclases, pyroxenes (orthopyroxenes and clinopyroxenes), olivines, accompanied with accessory phases such as metal oxides and other metal-bearing phases \cite{Lucey2006}.

\subsection{Areas of interest}

\begin{figure*}
    \centering
    \includegraphics[width=0.7\textwidth]{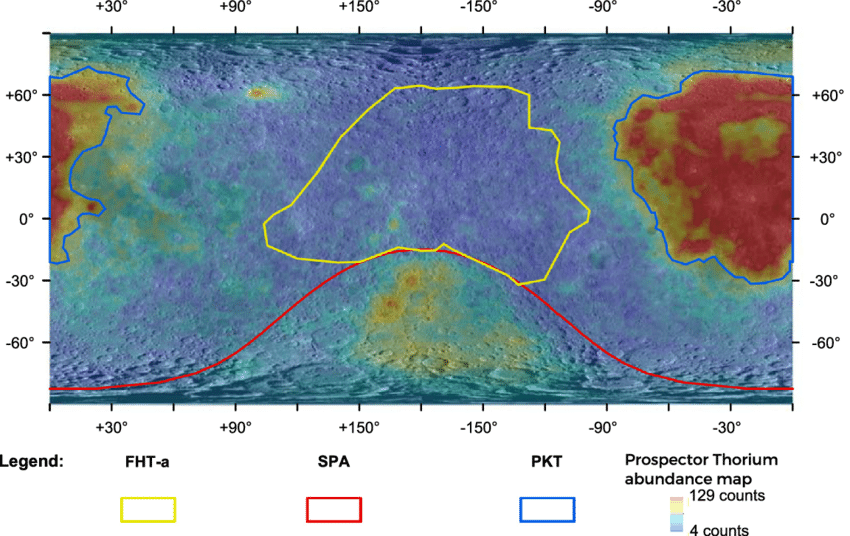}    
    \caption{Outline of the three major terranes as defined by \cite{Jolliff2000}, with the figure taken from \cite{MARTINOT2020113747}. The PKT is in blue, the SPAT is in red, and the FHT-a is in yellow. The surface that is not comprised in the PKT, SPAT, or FHT-a is the FHT-o.}
    \label{fig:major_terranes}
\end{figure*}

Three major terranes have been distinguished on the Moon: the Procellarum KREEP Terrane (PKT), the Feldspathic Highlands Terrane (FHT), and the South Pole-Aitken Terrane (SPAT) \cite{Jolliff2000}. A map of the major terranes of the Moon is in Figure \ref{fig:major_terranes}. The Procellarum KREEP Terrane (PKT) is a large oval-shaped terrain enclosing the lunar mare regions and highlands on the near side of the Moon. PKT is characterized by high abundances of potassium (K), rare-earth elements (REE), and phosphorus (P) \cite{Jolliff2000}. The Feldspathic Highlands Terrane (FHT) exhibits an anorthositic composition and is distinguished by low FeO and thorium levels. It is divided into two sub-terranes with distinct geochemical signatures: FHT-a, predominantly anorthositic, encompassing the majority of the lunar farside beyond the SPAT, and the outer FHT (FHT-o), which comprises cryptomare and ejecta from impact basins \cite{MARTINOT2020113747}. The South Pole-Aitken Terrane (SPAT), located on the far side of the Moon, represents a huge impact basin with anorthosites enriched with FeO (about 10 wt\%) when compared to the typical anorthositic crust on the Moon \cite{Lucey1998}. The impact exposed the deep mantle rocks, which are composed primarily of olivine and pyroxene, with elevated Ti, Fe, Th and K levels \cite{Moriarty2021} \cite{Moriarty2021Nature}. 

The Mare Tranquillitatis is particularly rich in \chem{TiO_2} \cite{Surkov2020}. Additionally, it has been found that the high titanium abundance may be due to ilmenite presence, as well as other titanium-bearing minerals (spinel group, rutile, or armalcolite) which were present in the Apollo mission samples in very low abundances. Ilmenite (\chem{FeTiO_3}) is known to be an excellent source of iron and titanium - these elements are critical in terms of building and maintaining lunar settlements.

\subsubsection{Mineral recognition by spectral absorption features}

Minerals can be identified by the molecular absorption bands in the near-infrared to the visible range of light \cite{Clark2019}. The most prominent absorption bands appear at approximately 1 $\mu$m and 2 $\mu$m, also commonly referred to as Band I and Band II. The absorption bands at $\lesssim$ 1 $\mu$m are caused by crystal field and charge transfer effects, which can occur when transition metals with filled d-orbitals reside in a ligand field \cite{Burns1993}. The strong absorption around 2 $\mu$m could also be attributed to crystal field transitions, particularly from Fe2+ (and other metals) occupying the M2 site, for instance, in pyroxenes. Additionally, at longer wavelengths ($\gtrsim$ 1 $\mu$m), the absorption could be caused by the characteristic molecular vibrations of cations, which can also provide information on the molecule's structure and composition. Each mineral, thanks to its unique elemental composition and crystal structure, has its own individual spectrum composed of individual sets of absorption peaks, based on which it can be identified. 
The 0.4 $\mu$m to 2.6 $\mu$m wavelength of the infrared light is a convenient range, as many rock-forming minerals have spectral characteristics in this span \cite{Adams1974} \cite{Lundeen2011}. In these wavelength ranges, the pyroxene absorptions are around $\sim$ 1 $\mu$m and $\sim$ 2 $\mu$m. Here, orthopyroxene can be found by the absorption bands centered at 0.9 $\mu$m and 1.8 $\mu$m, while clinopyroxenes demonstrate prominent absorption bands at 1.05 $\mu$m and 2.3 $\mu$m \cite{Adams1974}. Furthermore, the absorption at 1.2 $\mu$m relative to the absorption at the M2 site in pyroxene can be directly related to $\text{Fe}^{2+}$ in the M1 site \cite{Klima2007, Klima2008}. Interestingly, the 1 $\mu$m shifts regularly to longer wavelengths as a function of $\text{Ca}^{2+}$, as well as to some extent the M2 site at 2 $\mu$m \cite{Klima2011}. It has been found that useful relative relationships between composition and band center exist in most cases with pyroxene, but a universal quantitative relationship can not be determined without geological context \cite{Moriarty2016}. The plagioclase spectrum is classified mostly by its absorption between 1.1 $\mu$m and 1.3 $\mu$m and can be detected by flattening of the pyroxene reflectance “peak” between the 0.9 $\mu$m and 1.9 $\mu$m absorption bands if significant amounts of plagioclase are present \cite{Crown1987}. Since the amount of plagioclase has to be at least 90 \% for the absorption features to be distinguished at a location \cite{Barthez2023}, due to its absorption bands being masked by those of mafic minerals, it is also generally accepted that the lack of mafic absorption features in the near-infrared region is due to the presence of plagioclase \cite{Adler1977}. For olivines, Fe2+-endmember fayalite is characterized by absorption at 0.9, 1.08, and 1.3 $\mu$m, while Mg-endmember forsterite should show absorption at 0.84, 1.05, and 1.25 $\mu$m \cite{Horgan2014} \cite{RELAB:fayalite} \cite{RELAB:fosterite}. Ilmenite and related Ti and Ti-Fe oxides are recognizable in VNIR by two major absorption regions: one near 0.5 $\mu$m and a second broader absorption region in between $\sim$ 1.3 and 1.6 $\mu$m, distinctive from pyroxenes or olivines \cite{Izawa2021}. Ilmenite is traditionally difficult to identify from orbit \cite{besse2014volcanic}. However, in laboratory analysis, the presence of ilmenite strongly affects the overall continuum and pyroxene absorption band properties, especially for the fine particle size samples \cite{Moriarty2016}. Finally, minor but spectrally influential Mg-/Al-rich spinels might also affect observed spectra due to their extremely broad absorption region spanning $\sim$ 1.3 till $\sim$ 3 $\mu$m spectra \cite{Moriarty2023}.  

The Moon’s surface is inherently composed of both pure and mixtures of minerals \cite{Zhang2021}, hence there are very few monomineralic materials on the surface of the Moon \cite{Moriarty2016}. When multiple minerals are present, they may not be a linear combination of their spectral absorbance features. Moreover, the spatial resolution also affects the spectral features, since a lower resolution covers larger areas, hence multiple mineral types could contribute to the measurements. These factors result in difficulties when using unmixing techniques, and in this paper, the groups of clusters show different spectrally dominant mineralogies, albeit not necessarily pure “endmember” minerals. 

In addition, various other effects may influence the petrologic interpretation of the lunar remote sensing observations. For instance, the younger more recently space weather exposed mare surfaces should also demonstrate less modified spectral features, than mature highland soils \cite{Mouelic2000} \cite{Staid2000}. The topographic shading correction also presents a challenge for the analysis of the remote sensing measurements, in particular at high latitudes \cite{Bussey2010}.    

\subsubsection{Mineral maps of the Moon}

Several global lunar mineral maps have been created of the abundances of lunar minerals, among others: Ilmenite has been mapped with the Moon Mineral Mapper (M3) data \cite{Surkov2020} and in \cite{Bhatt2019} Ti, Fe, Mg, and Ca estimated using multivariate regression and ML methods; \chem{TiO_2}, which is a component of ilmenite, was mapped by The Lunar Reconnaissance Orbiter Camera (LROC) Wide Angle Camera (WAC) \cite{Sato2017}; with empirically derived spectral parameters plagioclase, orthopyroxene, clinopyroxene, olivine, and FeO are mapped from absolute latitudes less than 50 degrees with Kaguya data in \cite{Lemelin2019}, and on the poles in \cite{Lemelin2022}; with Clementine, the FeO and titanium abundance was estimated \cite{Lucey2000_1}, and a color-ratio map was made \cite{Lucey2000}; with Lunar Prospector gamma rays and neutron spectrometer maps were made of Ti, Fe, Th, K \cite{Lawrence1998} \cite{Elphic1998}. 

An algorithm for estimating the Fe abundance based on the properties of the pyroxene-related absorption band near 2 $\mu$m has been introduced by \cite{Megha_Upendra_Bhatt2012}, using hyperspectral image data of M3. In \cite{Bernhardt2017} a genetic algorithm was used to perform spectral unmixing and estimate plagioclase, pyroxene, and olivine weight percentages. In \cite{Qiu2022} FeO and \chem{TiO_2} were estimated by training a convolutional neural network to construct an optimized spectral inversion. 

\subsection{Autoencoders for dimensionality reduction and clustering}

According to \cite{Aggarwal2001}, as the number of dimensions in a dataset increases, the ratio of the distance of a data point to its nearest and farthest neighbors approaches 1, meaning all pairs of points are almost equidistant from each other, which can negatively impact the performance of clustering algorithms that aim to model the differences between the data points. Fortunately, several dimension reduction techniques are effective in preserving the characteristic information of the data while increasing the contrast of the point separations, thus facilitating the clustering of high-dimensional data. Autoencoders are among such dimensionality reduction, data compressing, and unsupervised deep learning techniques. In particular, variational autoencoders have been successfully used to effectively represent input data distributions in a low-dimensional abstract compressed space, also named the latent space. The latent space captures the major underlying structure and relationships of the input data and allows reconstruction of the feature-learned data in the original dimension, thus  (“decoding”) from the latent space. Recent studies have shown the usefulness of variational autoencoders for clustering high-dimensional data \cite{Lim2020}, and in some cases, convolutional autoencoders have also been used \cite{Nellas2021}. Studies have shown that deep autoencoders, in combination with k-means unsupervised learning, can provide an efficient clustering algorithm \cite{lu2021dac}. Moreover, autoencoders have shown to be useful in denoising data since the decoder learns to ignore noise in the input data and focus on the underlying structure \cite{NIPS2012_6cdd60ea, poole2014analyzing, vincent2010stacked}. Because of the success of both variational autoencoders and convolutional networks, a convolutional variational autoencoder was developed and implemented in the analysis presented in this paper.

\section{Data}

The Moon Mineral Mapper (M3) from the Chandrayaan-1 has a wide range in the spectral dimension, which is only surpassed by the Imaging Infrared Spectrometer (IIRS) from the Chandrayaan-2 mission regarding data of the Moon. M3 is a spectrometer that operates in the solar-dominated portion of the electromagnetic spectrum with wavelengths from 0.43 $\mu$m to 3.0 $\mu$m \cite{Green2011} \cite{Pieters2009}. The M3 was chosen for this analysis since it has a wide range of spectral wavelengths and a good, although not perfect, coverage of the Moon \cite{Boardman2011}. The IIRS also has a wide range in spectral regions (0.8 - 5.0 $\mu$m \cite{Chowdhury2020}), but at the time of writing, the corrected and calibrated data has not been fully released. 

The published M3 reflectance data and photometrically corrected \cite{Besse2013} were used, but not including wavelengths over 2.5 $\mu$m, since additional corrections would be needed to correct for thermal additions \cite{Li2016}. 

\begin{figure*}
    \centering
    \includegraphics[width=1.\textwidth]{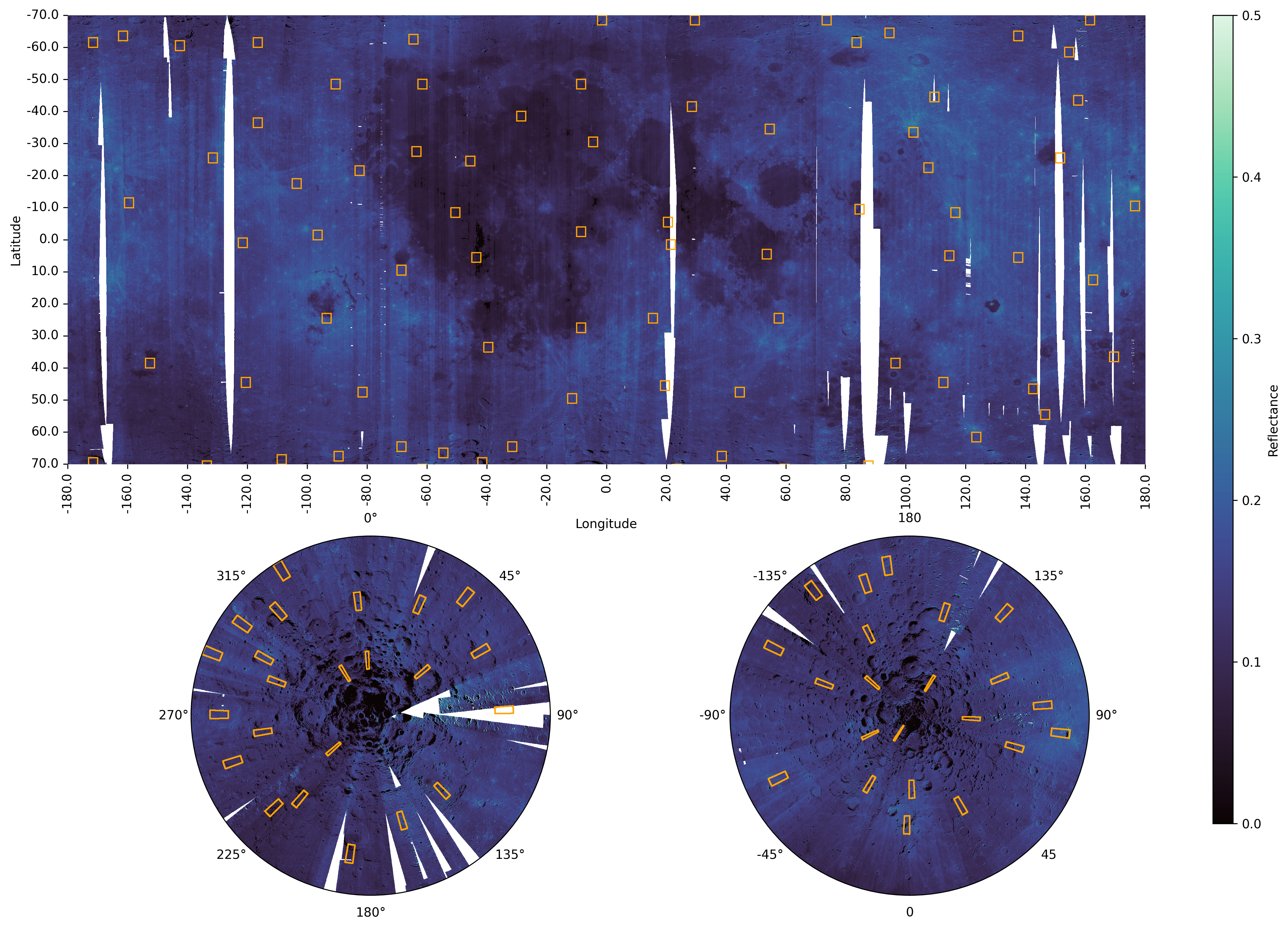}
    \caption{The Moon Mineralogy Mapper (M3) reflection data of the Moon. The orange squares highlight the areas of the dataset used to train the convolutional variational autoencoder. The upper map is displayed in an equidistant cylindrical projection with a maximum latitude of 75 degrees. The bottom left and right show the Moon's south and north poles, respectively, in a stereographic projection with a minimum of 60 degrees.
    \label{fig:training_data}
    }
\end{figure*}

The training data was selected by hand for the first 20 images, with a particular interest in high weight percentages and different mineral selections as seen in Figure \ref{fig:training_data}, and 80 areas were chosen at random. Each area is 3 degrees wide in both latitude and longitude. At the resolution of 0.05 degrees, the total spectral dataset used for training, validation, and testing is $100 \text{ AOI} \times \frac{3^{\circ}}{0.05^{\circ}} \times \frac{3^{\circ}}{0.05^{\circ}} = 360000$ spectral lines, where AOI is Area Of Interest.

\subsection{Data selection and pre-processing}

The data was selected by a maximum observation angle of 180 degrees, a maximum sensor-to-sun zenith of 90 degrees (i.e., emission angle), and a maximum sensor zenith of 25 degrees (i.e., incidence angle). The choice of angles is intended to optimize the data coverage while minimizing the cuts on the angles to obtain the best data quality. Since the M3 dataset has high observation angles, their cuts are loose in order to obtain good coverage of the Moon. The data periods of M3 varied by having experimental effects, which were mostly thermal, as the orbit was changed from 100 km to 200 km during the data-taking period. This analysis selects the data using a data-driven approach, meaning that all periods of M3 observations are included. However, the spectral data is removed automatically by the quality cuts on an image-by-image basis. The M3 instrument recorded data in a global mode and a target mode with a higher resolution. This analysis only uses the global mode dataset. Furthermore to the orbital cuts, spectra with an average spectrum of less than 0.05 reflectance are neglected. These values correspond to pixels with zeros (permanently shadowed regions, detector effects, or missing data). Outliers are removed on an image-by-image basis, where values that are ten standard deviations away from the average are considered outliers. The deviations removed by this cut are mostly due to thermal effects, as the infrared readings deviate from the average values and are outliers. 

After the data selection, each individual image from M3 is rebinned to 0.05 by 0.05 degrees. The highest resolution was chosen based on coverage on the poles, as coverage of the poles was a priority. The pixels are then averaged over all periods, such that the fluctuations are minimized, and data quality overall improves by the average. The combination of overlapping spectral images provides more robust spectral absorptions. The rebinning to a resolution of 0.05 degrees results in approximately a pixel width of 1.5 km - 2 km, depending on the position on the Moon.

\section{Methods}

\begin{figure*}
    \centering
    \includegraphics[width=\textwidth]{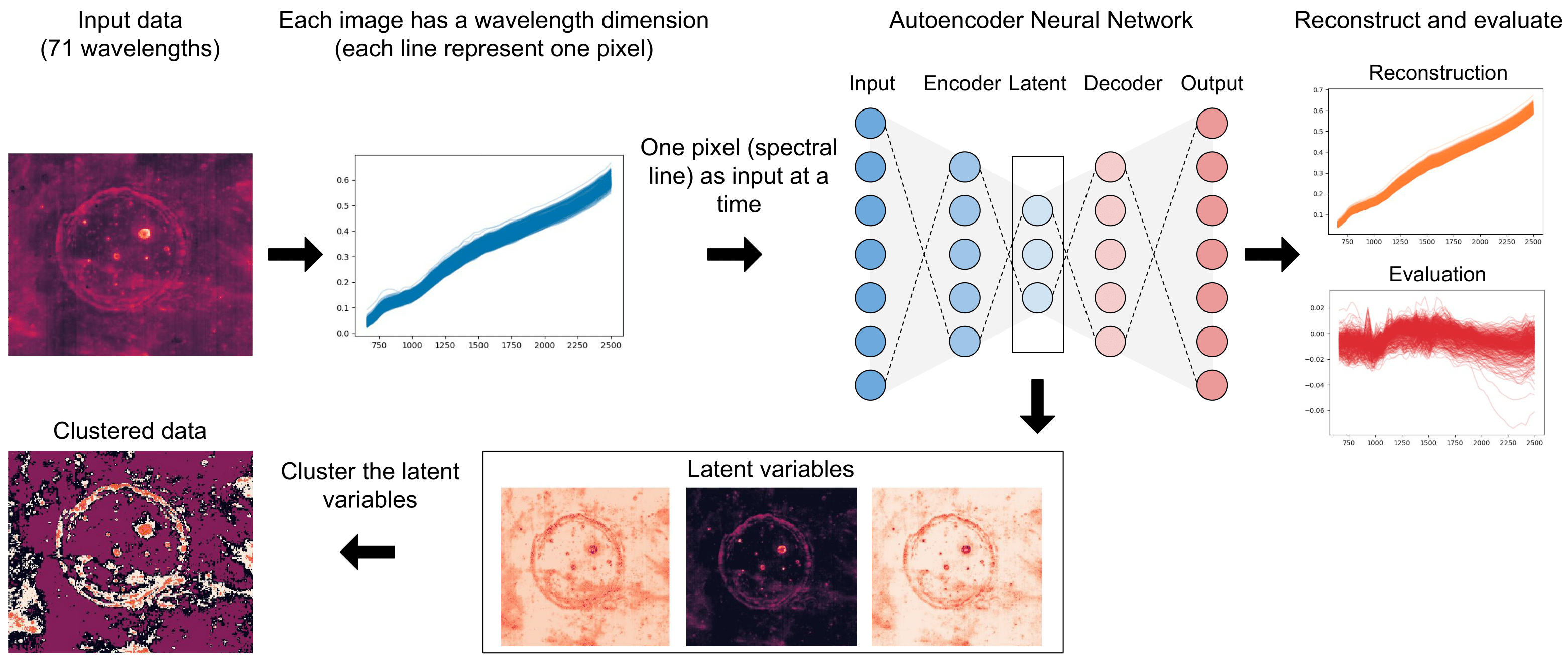}
    \caption{Overview of the process of training the neural network and clustering of data. The spectra are passed through a symmetric convolutional bottleneck autoencoder and reconstructed. The loss between the original spectra and the reconstructed spectra is minimized. The encoded latent variables contain the relevant features needed to reconstruct the spectra. Therefore, the latent variables are used to represent the spectra in low-dimensional space. After the latent variables are obtained for each pixel, they are clustered by k-means.
    }
    \label{fig:methods}
\end{figure*}

The aim of feature extraction is to find a mapping from high-dimensional space to low-dimensional space that reduces redundant information and preserves important information. The overall methodology is depicted in Figure \ref{fig:methods}. The process first passes the spectral lines through a bottleneck neural network to significantly reduce data dimensionality. The compressed representation of the data, which are the latent variables, is then clustered with the k-means unsupervised ML algorithm to produce a clustered image. Detailed explanations of the neural network and clustering approach are provided in the subsequent sections.

\subsection{Variational Autoencoders}

The autoencoder is an unsupervised learning approach where the network encodes a dataset $\mathbf{x}$ into low-dimensional representations and then reconstructs the dataset through a decoder \cite{Rumelhart1988}. In autocoders, the architecture can be decomposed into two steps: encode the dataset $\mathbf{x}$ into a lower-dimensional representation (latent variables), $\mathbf{z}$ using a neural network $g_\phi: \mathbf{x} \rightarrow \mathbf{z}$, and decode the representation $\mathbf{z}$ into a reconstruction $\hat{\mathbf{x}} = h_\phi(\mathbf{z})$. 

The variational autoencoder extends this concept by introducing probabilistic regulation of the latent space. Here, the latent variable $\mathbf{z}$ captures generative factors of $\mathbf{x}$, with a chosen prior distribution, typically a Gaussian distribution (e.g., $\mathcal{N}(0 , I)$). Variational Inference (VI) overcomes the intractability of the true posterior $p_\theta(\mathbf{z} | \mathbf{x})$ by introducing an approximate posterior, denoted as $q_\phi(\mathbf{z} | \mathbf{x})$, typically chosen from a variational family, often the Gaussian family.

The introduction of this approximate posterior allows the likelihood $p_\theta (\mathbf{x})$ to be expressed as an expectation over $q_\phi(\mathbf{z} | \mathbf{x})$. To ensure numerical stability, we utilize log-space computation, resulting in the Evidence Lower Bound (ELBO) $\mathcal{L} (\mathbf{x})$ \cite{Kingma2022}. The ELBO is defined as:

\begin{eqnarray}
\mathcal{L}(\mathbf{x}) :=& \mathbb{E}_{q_\phi(\mathbf{z} | \mathbf{x})} \left[ 
\log p_\theta(\mathbf{x} | \mathbf{z}) - \log q_\phi(\mathbf{z} | \mathbf{x}) + \log p_\theta(\mathbf{z}) 
\right] \\ =&
\overbrace{
\mathbb{E}_{q_\phi(\mathbf{z} | \mathbf{x})} \left[ \log p_\theta(\mathbf{x} | \mathbf{z})\right]
}^{\text{Reconstruction Error}}
- 
\overbrace{
\mathcal{D}_{\operatorname{KL}}\left(q_\phi(\mathbf{z}|\mathbf{x})\ |\ p(\mathbf{z})\right)
}^{\text{Regularization}}
\end{eqnarray}

\begin{figure*}
    \centering
    \includegraphics[width=0.4\textwidth]{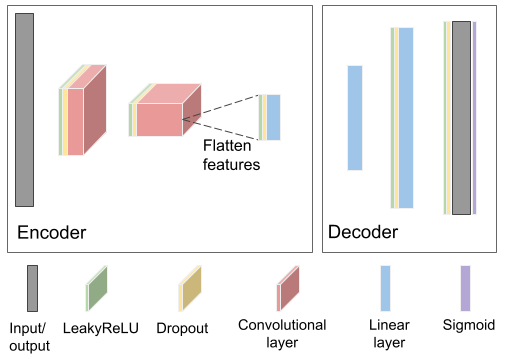}
    \caption{Structure of the variational autoencoder. The input has the form of the number of wavelengths. The convolutional layers decrease the features of the data while increasing the number of kernels in the convolutional layers. E.g. in the encoder, the data is more compressed going from left to right. After the convolutional layers, the features are flattened before going into the bottleneck consisting of a linear layer. The decoder is the inverse of the encoder, consisting of linear layers, with the sigmoid activation function at last.}
    \label{fig:neural_network}
\end{figure*}

In this paper, the neural network architecture is adapted from \cite{Higgins2016}, with a variation of having a convolutional bottleneck structure in the variational autoencoder. Batch normalization layers were introduced after each convolutional layer, where the batch normalization layers normalize the input values based on the mean and variance of the mini-batch during training. The batch normalization layers can enhance training efficiency by reducing internal covariate shifts and allowing each layer to learn more independently \cite{DBLP:journals/corr/IoffeS15}. The neural network structure is in Figure \ref{fig:neural_network}, and the dimensions and layer types are listed in Table \ref{tab:neural_network}.

This analysis used a simple Mean Standard Error (MSE) loss function to calculate the loss. The algorithm is trained and tested with three different datasets: Training (80 \%), validation (10 \%), and test dataset (10 \%). 

\begin{table}[H]
\centering
\begin{tabular}{ccc} 
\toprule
\textbf{Layer \#} & \textbf{Layer Type} & \textbf{\# Output Units}\\
\midrule
1 & Input (linear) & $N$ \\
2 & Conv1d & $N/2 \times $( 16 channels) \\
3 & Conv1d & $N/4 \times $(32 channels)\\
4 & Linear& L\\
5 & Linear & 32\\
6 & Linear & 64\\
7 & Linear& $N$\\
\bottomrule
\end{tabular}
\caption{Dimensions and layers of the encoder/decoder architecture. The decoder contains the same layers but inversely. $N$ is the size of the input, and $L$ is the size of the latent variables. In this paper, $N=71$, kernel size $=3$, and $L=5$. The convolutional layers have the channel dimension, denoted ($\times$), and the linear layers are 1-dimensional.}
\label{tab:neural_network}
\end{table}

\subsection{Clustering Approach}

The optimal number of clusters was determined by the Silhouette \cite{Rousseeuw1987} and Davies-Bouldin \cite{Davies1979} metrics. The Davies-Bouldin index is calculated as the average similarity of each cluster with a cluster most similar to it, where a high score means a higher similarity and low separation between the clusters. The Silhouette method is a measure of how similar data is to its own cluster compared to other clusters. Its values are between -1 and 1, where a value of 1 is well-separated clusters, and a value below 0 is data points in the wrong cluster or a nonoptimal number of clusters. Both scores were calculated in the range of clusters between 2 and 9. The minimum value, and hence the best choice of the number of clusters, of the Davies-Bouldin index is at 5 clusters, with a Davies-Bouldin index of 0.49. The Silhouette score had its maximum, and therefore best choice of number of clusters, also at 5, with a value of 0.61. Therefore, the resulting choice of clusters is 5. 

It is important to note that in this analysis, there is no inherently “wrong” choice of clusters. The decision to use 5 clusters results in the most distinctive clustering, but alternative arguments, such as selecting two clusters to represent different characteristics (e.g., lunar mare and highlands regions), can also be considered valid.

The cluster centers were calculated using only the central latitudes ($|$latitude$| < 70$) for cluster center definition due to the inherent uncertainty associated with polar regions. Signals from these regions were collected under suboptimal conditions, with a large observation angle that could impact accuracy \cite{Besse2013}. 

\section{Results}

We present here the clustering of the latent variables results in the global Moon spectral cluster map. The spectra of the clusters of the Moon spectral cluster map are examined through reflectance and continuum-removed spectra. The mapping of the clusters is also presented and compared to results from Kaguya.

\subsection{Moon spectral cluster map}

\begin{figure*}
    \centering
    \includegraphics[width=1.\textwidth]{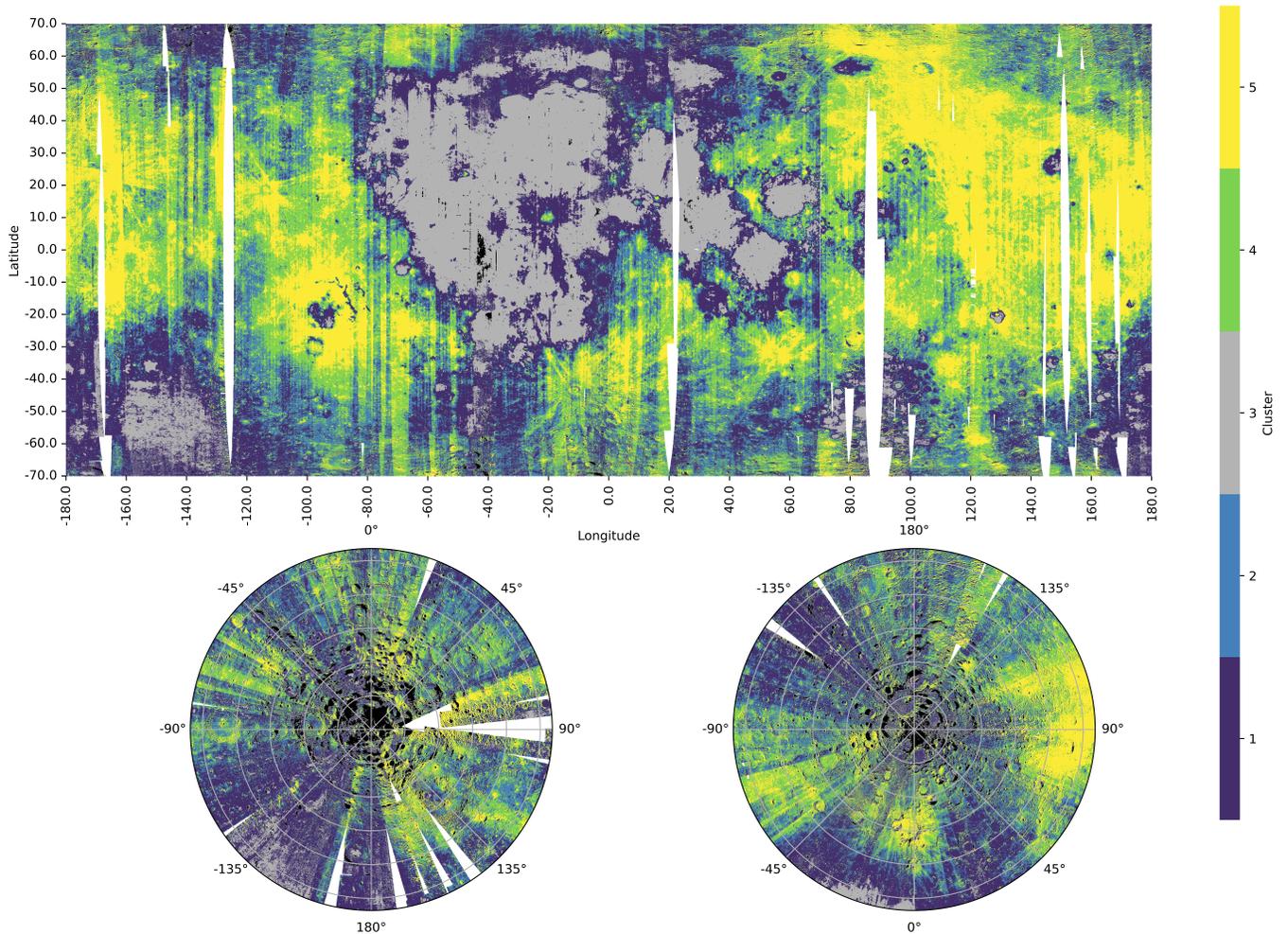}    
    \caption{The resulting Moon spectral cluster map from the deep learning clustering approach. The top is in equirectangular projection from -70 to 70 degrees in latitude. The bottom left is the south pole in stereographic projection from -60 degrees latitude, and the bottom right is the north pole in stereographic projection from 60 degrees latitude. The black regions have zero reflectance, and the white regions have no data. }
    \label{fig:mineral_map}
\end{figure*}

The Moon spectral cluster map is in Figure \ref{fig:mineral_map}. Interestingly, cluster 3, which is prominent in Figure \ref{fig:mineral_map}, is found mostly in the maria and very sparsely at the poles. Cluster 3 also covers parts of the South Pole-Aitken Terrane (SPAT), which is high in clinopyroxene but is not a Mare basalt \cite{Moriarty2018} \cite{Moriarty2015}. Cluster 1 encapsulates the Moon’s mare region consisting of olivine, FeO, orthopyroxene, and clinopyroxene. In addition to these established terranes, the highlands surrounding the Imbrium basin warrant further investigation. These highlands, rich in mafic minerals such as those found in the noritic anomaly identified by \cite{isaacson2009northern} and the mafic highlands (MH) described by \cite{wu2018geology}, may represent a distinct terrane. The mafic composition suggests that these highlands are composed of exposed lower crustal material, excavated during the Imbrium basin impact. Interestingly, Cluster 1 identified in this study closely corresponds to these highlands.

There is a slight polarity in the locations of the spectral clusters in the Moon spectral cluster map. For example, cluster 1 is most prominent at the poles, however, it is also found near the equator, surrounding the mare region. Previous analysis showed a high polar dependency of OH with the Moon Mineral Mapper \cite{Li2017}. However, with a cut-off of 2.5 $\mu$m in this analysis, the absorption bands for \chem{H_2O}/OH are not prominent, and therefore the contributions from \chem{H_2O}/OH can not be implied from these clustering results. Clusters 4 and 5 are present on the Moon in regions dominated by plagioclase \cite{Ohtake2009}. Cluster 1 is in other areas dominated by mostly iron and orthopyroxene but also clinopyroxene \cite{Moriarty2018}. Cluster 2 is in the transition areas by clusters 1 and 4. 

\subsection{Spectral classifications}
\begin{figure}
    \centering
    \includegraphics[width=0.48\textwidth]{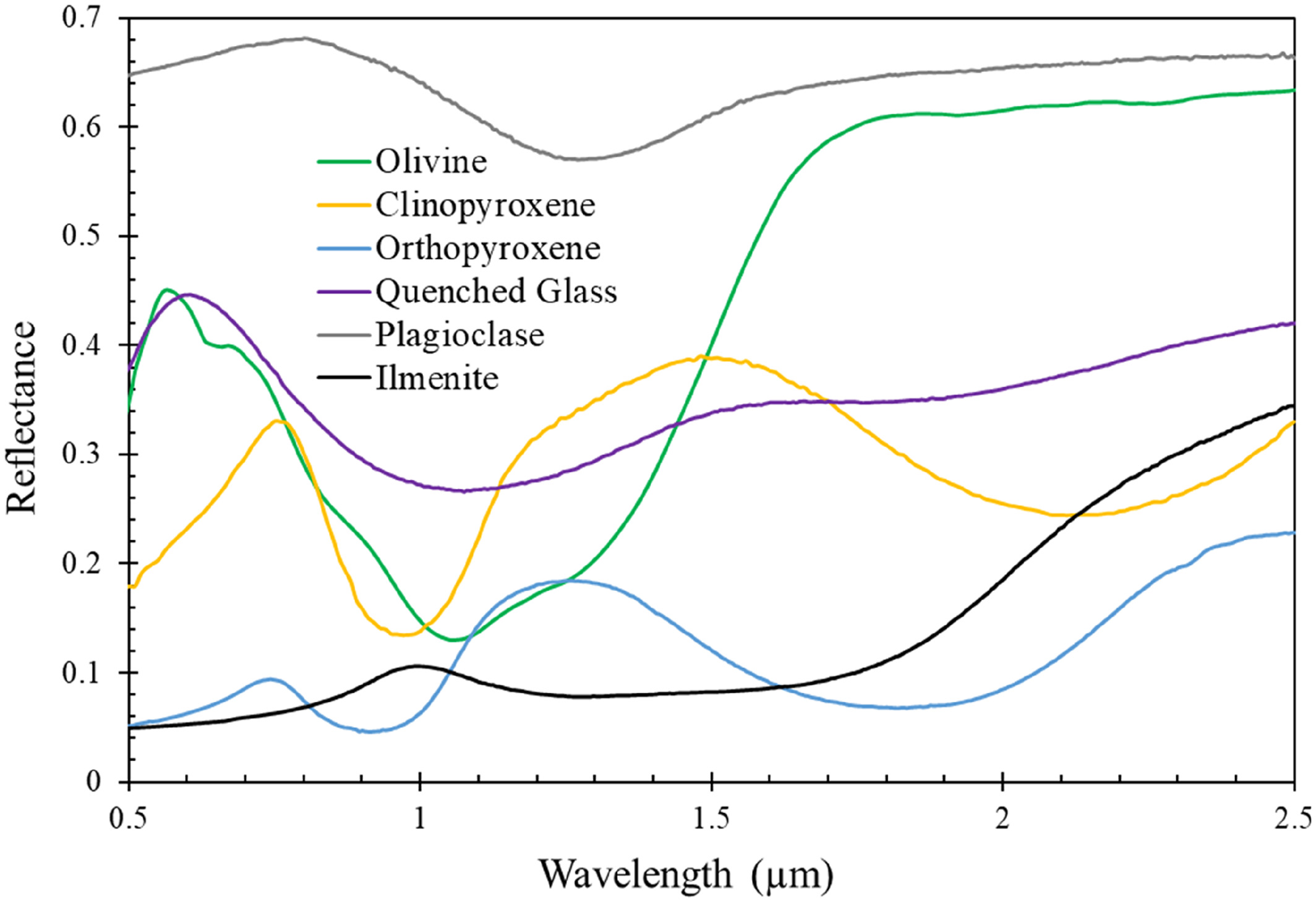}    
    \caption{Representative reflectance spectra of selected minerals from \cite{Zhang2021}. The samples used are from the RELAB database and are the following: Ol (PO-CMP-026), Cpx (LS-CMP-009), Opx (PP-RGB-080), glass (LS-CMP-035-A), Pl (LR-CMP-217), and Ilm (LR-CMP-218). Cpx, clinopyroxene; Ilm, ilmenite; Ol, olivine; Opx, orthopyroxene; Pl, plagioclase.
    }
    \label{fig:spectra_db}
\end{figure}

\begin{figure*}
    \centering
    \includegraphics[width=.8\textwidth]{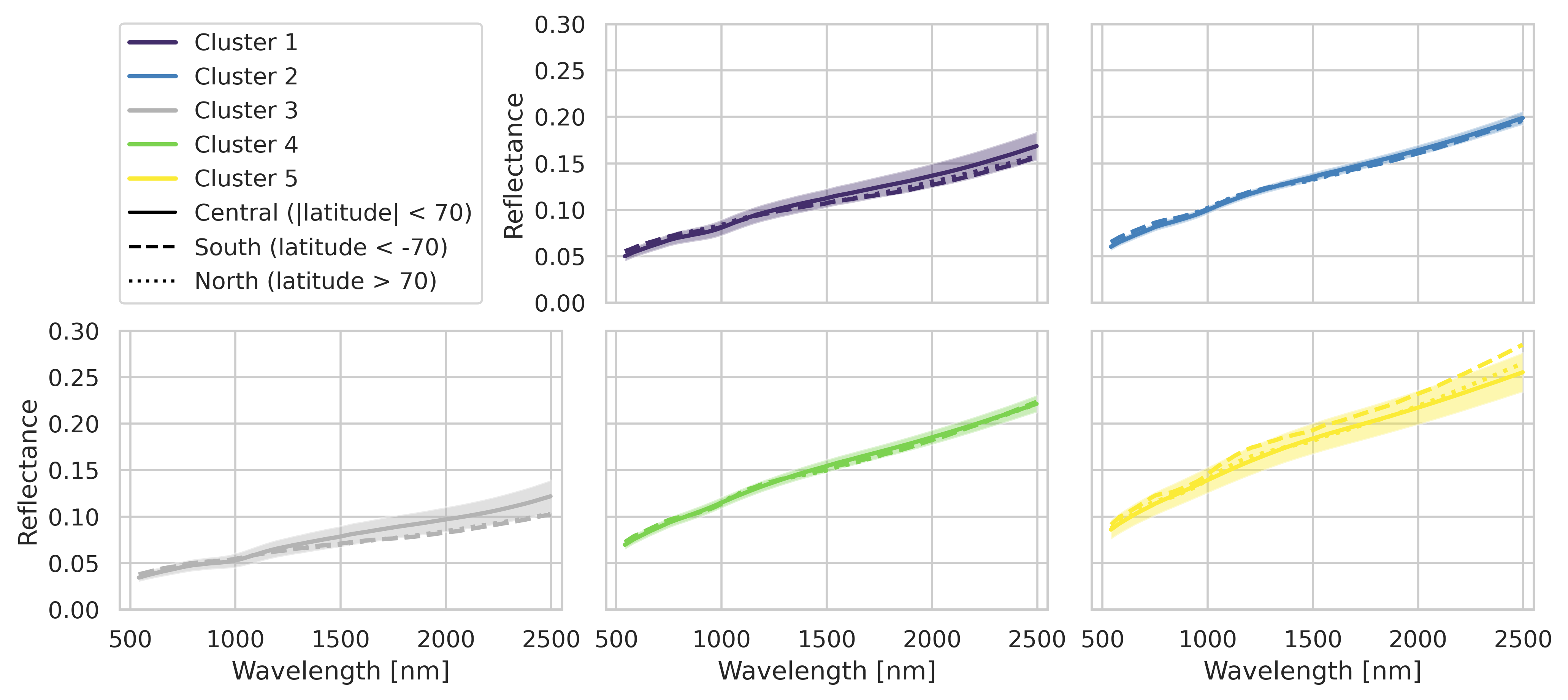}    
    \caption{Average reflectance spectra from the 5 clusters. The average continuum removed spectra are calculated for the central, north, and south latitudes of the Moon. The shaded areas are one standard deviation from the average.
    }
    \label{fig:spectra_clusters}
\end{figure*}

\begin{figure*}
    \centering
    \includegraphics[width=0.8\textwidth]{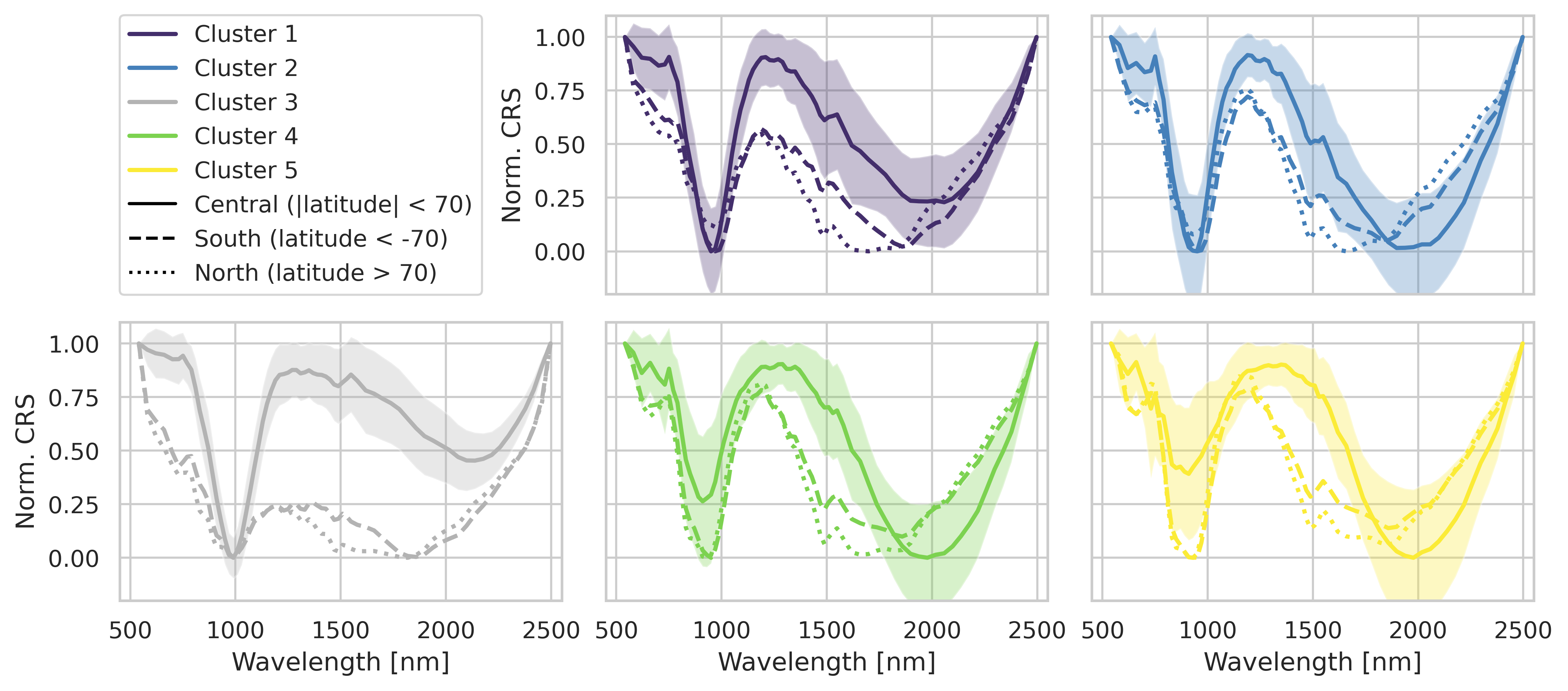}    
    \caption{ Average normalized continuum-removed spectra from the 5 clusters by the convex hull method. The average continuum removed spectra are calculated for the central, north, and south latitudes of the Moon. The shaded areas are one standard deviation from the average.
    }
    \label{fig:spectra_clusters_crs}
\end{figure*}

The reflectance spectra of a few selected minerals known to be on the Moon are in Figure \ref{fig:spectra_db}, which can be used to study the spectra of the 5 clusters. In Figure \ref{fig:spectra_clusters} the reflectance from the 5 clusters is shown, and in Figure \ref{fig:spectra_clusters_crs} the normalized Continuum Removed Spectra (CRS) for each cluster is shown. Only the spectra with a standard deviation below 0.25 (with the spectra normalized) are used for the average, since otherwise the noise contribution is too large for a viable continuum-removed signal. There are differences in not only the magnitude of the reflectances between the 5 clusters but also in the absorption wavelengths. From Figure \ref{fig:spectra_clusters}, it is clear that cluster 3, corresponding to the maria, has the lowest reflectance. The south pole’s spectra, which is also in cluster 3, show an even lower reflectance. Since the Moon’s rotation pole to the ecliptic plane has a 1.54 $\degree$ inclination, the Sun never rises more than a few degrees of elevation angle above the horizon at the lunar polar regions \cite{bryant2009lunar}. Therefore, the data at the poles can have a lower signal-to-noise ratio due to the sunlight conditions, even after corrections were applied to the data \cite{Besse2013}. The absorption features, i.e., the minima of the continuum removed spectra from Figure \ref{fig:spectra_clusters_crs}, are also relatively smaller for the polar regions. This means that the absorptions are less prominent at the poles, most likely due to the lower observation angles, lower data quality due to higher sun beta angles,  permanently shadowed regions, or other reasons that can not be reliably determined. 

In  Figure \ref{fig:spectra_clusters_crs}, it is evident that the relative minima of Band I and Band II (at ~1 and ~2 $\mu$m) vary depending on the cluster. For cluster 1, the average CRS has a deeper minimum at Band I than at Band II. The strong absorption at Band I  can be due to pyroxenes absorptions, i.e., orthopyroxene (at 0.9 $\mu$m) and clinopyroxene (at 1.05 $\mu$m). Cluster 2 shows a similar shape to cluster 1, although it is important to notice that the absorption at Band II is stronger, as in the case of cluster 1. Recent studies have found that minor spinel components in highlands soils may also affect reflectance spectra at $\gtrsim$ 2 $\mu$m due to their strong and broad absorption at these wavelengths \cite{Moriarty2023}, which could possibly contribute to the prominent absorption feature apparent at $\gtrsim$ 2$\mu$m for cluster 2. Remarkably, cluster 3 shows a strong absorption at 1 $\mu$m, with a shift compared to clusters 1 and 2. This shift can arguably be due to the contribution of olivine, which has the most dominant absorption around 1.05 $\mu$m. Clusters 4 and 5 show a greater relative absorption at 2 $\mu$m with respect to the absorption at 1 $\mu$m, compared to the other clusters. In particular, cluster 5 has a relatively larger absorption at 2 $\mu$m than at 1 $\mu$m. The shift of the dominant absorptions can be largely due to a contribution by plagioclase, which can result in a flattening effect at the pyroxene reflectance “peak” between 0.9 $\mu$m and 1.9  $\mu$m \cite{Crown1987}. 

\begin{figure*}
    \centering
    \includegraphics[width=.8\textwidth]{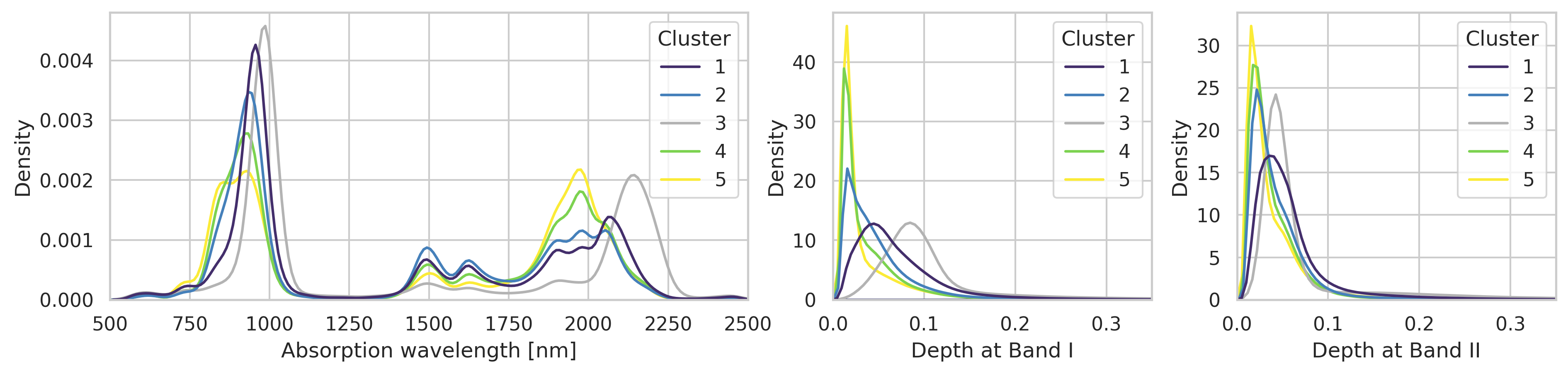}    
    \caption{Densities of the two most dominant measured features from the continuum removed spectra per cluster. From left to right is the absorption wavelength, depth of Band I, and depth of Band II.
    }
    \label{fig:absorption}
\end{figure*}

Figure \ref{fig:absorption} shows the densities of absorption wavelengths and depth of the absorption features at Band I and Band II for each cluster. Band I is the feature with the largest absorption range of 0.75 $\mu$m - 1.25 $\mu$m, and Band II in the range of 1.25 $\mu$m - 2.5 $\mu$m. Only features with an absorption depth of more than 0.01 are considered in order to exclude variations in the data due to noise. A cut-off on absorption depth is essential to filter out minor fluctuations and avoid considering them as genuine absorptions. However, this also means that featureless spectra with absorption less than 1\% are not visualized in Figure \ref{fig:spectra_clusters_crs} and \ref{fig:absorption}. This predominantly affects cluster 5, which contains the primarily featureless data. As shown in Figure \ref{fig:absorption}, absorption depth tends toward 0 (with a cut-off at 0.01) for cluster 5. The absorption depths are larger for clusters 1, 2, and 3, which relates to the presence of pyroxenes generally having a significant impact on the absorption depths. On the other hand, clusters 4 and 5 show very shallow absorption depths, possibly due to the flattening effect from the presence of plagioclase and an absent or small contribution of pyroxenes. The absorption wavelengths for the 5 clusters clearly show that the clusters have a shift in absorption wavelengths by Band I, e.g., cluster 3 has the highest absorption wavelength in Band I of an average of 963 nm. Interestingly, cluster 1 exhibits a strong absorption feature centered around 932 nm (Table \ref{tab:absorption}, indicative of orthopyroxene (opx). This observation further supports the unique nature of these highlands and their potential significance in understanding lunar geology. Further statistical details regarding the mean values of the Band I and Band II centers are presented in Table \ref{tab:absorption}.

\begin{table*}
\centering
\begin{tabular}{|l|c|c|c|c|}
\hline
\textbf{Cluster} & 
\multicolumn{2}{c|}{\textbf{Mean Absorption Wavelength [nm]}} & 
\multicolumn{2}{c|}{\textbf{Mean Depth}} \\
\cline{2-5}
 & \begin{tabular}[c]{@{}c@{}}Band I \\ (500 nm - 1250 nm)\end{tabular} & \begin{tabular}[c]{@{}c@{}}Band II \\ (1250 nm - 2500 nm)\end{tabular} & \begin{tabular}[c]{@{}c@{}}Band I \\ (500 nm - 1250 nm)\end{tabular} & \begin{tabular}[c]{@{}c@{}}Band II \\ (1250 nm - 2500 nm)\end{tabular} \\ 
\hline
1 & 932.08 & 1874.77 & 0.071 & 0.060 \\ \hline
2 & 914.82 & 1834.52 & 0.050 & 0.042 \\ \hline
3 & 963.33 & 2027.45 & 0.107 & 0.077 \\ \hline
4 & 903.12 & 1882.15 & 0.044 & 0.034 \\ \hline
5 & 891.42 & 1900.22 & 0.043 & 0.033 \\ \hline
\end{tabular}
\caption{The averages of the location of the most dominant absorption features of the spectra from the 5 clusters in Band I and Band II.}
\label{tab:absorption} 
\end{table*}

\subsection{Comparison to published maps}
\begin{figure*}
        \centering
        \includegraphics[width=1.\textwidth]{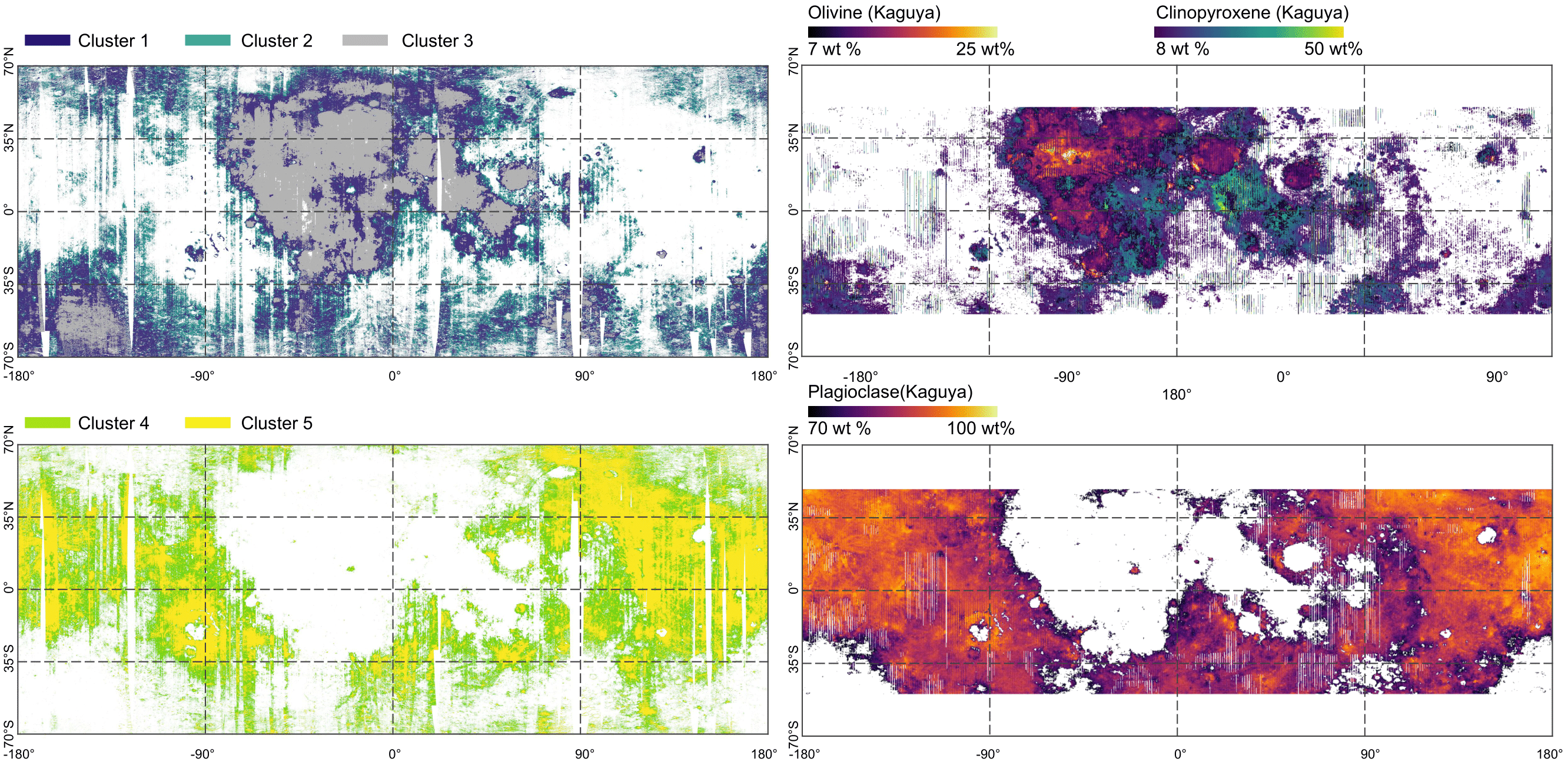}    
        \caption{Comparison of the 5 clusters and data from Kaguya \cite{Lemelin2019}, with applied masks of weight percentages in order to get a view of the strong absorptions. 
        }
        \label{fig:kaguya_comparison}
    \end{figure*}

The primary mineralogy of the clusters was determined by comparison to the minerals maps from Kaguya \cite{Lemelin2019}. The approach used in \cite{Lemelin2019} was based on Hapke’s radiative transfer equations to model the reflectance spectra of mineral mixtures of plagioclase, olivine, and clinopyroxene, among others. The paper used a continuum removal and a gradient descent algorithm to estimate the mineral abundances of the central peaks. A selection of the mineral abundance results from Kaguya is in Figure \ref{fig:kaguya_comparison}. The resolution of the Kaguya results was about 60 meters/pixel. Notice that the ranges of the weight percentages of the Kaguya results have been adjusted to a higher minimum weight percentage, compared to the published results, in order to compare the most dominant mineralogies. 

The figure shows that there is a general agreement with the placements of plagioclase dominant regions and clusters 4 and 5, and the olivine and clinopyroxene dominant regions nicely overlap with clusters 1, 2, and 3. The comparison to the Kaguya plagioclase map reveals that cluster 5 covers the higher wt\% areas with plagioclase than cluster 4. Cluster 2 almost surrounds cluster 1 and 2, and overlaps mainly with the regions of the Kaguya results where the clinopyroxene weight percentages are low (around 8 \%), thereby describing a transition area where the mineralogy is changing to a more plagioclase dominant region, compared to clusters 1 and 2. The comparison to the Kaguya results of olivine, clinopyroxene, and plagioclase shows that the mineralogies of the lunar surface with dominant spectral features relate to the 5 clusters from the deep learning approach.

Comparing Figure \ref{fig:mineral_map} and \ref{fig:major_terranes} reveals the relationship between clusters and lunar major terranes. Clusters 1 and 2 predominantly coincide with the PKT area, and there is also a significant overlap between clusters 1 and 2 with the SPA region. This suggests a similarity in mineralogy between SPA and PKT in the visible-near-infrared range (M3 data). Lastly, FHT-a and FHT-o areas cannot be distinguished using the clusters from this study, as both are mainly characterized by clusters 4 and 5. 

While these clusters do not directly map to specific minerals like those in the Kaguya maps, they nonetheless indicate distinct combinations of mineral assemblages, geological characteristics, and space weathering effects across the lunar surface.

\section{Conclusion}
An unsupervised learning algorithm was introduced to cluster spectra from the Moon Mineral Mapper (M3). The research acts as proof of concept for a new method for hyperspectral analysis of the Moon images, which can be used for areas of interest or the Moon spectral cluster map, as shown in this paper. The method uses a convolutional variational autoencoder to reduce the dimensionality of the spectral data and extract relevant features. Then, a k-means algorithm is applied to cluster the latent variables into five distinct groups corresponding to different mineral compositions. The clustering algorithm uses no information about location or neighboring pixels; hence, the structures found are purely from the spectral data. 

The resulting Moon spectral cluster map shows the location of the 5 clusters. The mineralogy of the 5 clusters was described, their spectra were analyzed, and comparisons were made to published maps. The comparisons included the locations of pyroxenes, plagioclase, olivine, and major terranes of the Moon. The clusters correspond to mineral features on the Moon, and by quantifying the clusters, they are compared to mineral maps from Kaguya.

One of the strengths of the unsupervised clustering analysis is that it does not rely on any assumptions about the mineralogy of the Moon and does not contain any human bias. The number of clusters is chosen on an unsupervised basis, and the division and separation of clusters are also determined with an unsupervised approach. The encoder-decoder structure enables the model to learn deep features and classify the pixels to corresponding (pseudo-)labels based on these features, which could reveal hidden structural patterns in measurements missed in the supervised classification approach. Furthermore, due to the efforts it brings to understand the main spectral features of the Moon, it could also be used in future analysis as a form of validation tool in that it contains no assumptions on the data.

While our method has proven effective in providing a comprehensive understanding of lunar mineralogical distribution, we acknowledge its limitations. The Moon spectral cluster map, based on M3 data, describes the overall spectral features on the Moon but can not be used to estimate weight percentages. We propose further improvements and analysis, such as incorporating spatial information and merging data from multiple instruments, to enhance the accuracy and applicability of our method. 

In summary, our analysis using clustering methods provides a comprehensive understanding of lunar mineralogical distribution. The distinct clusters comprise various mineralogical compositions across the Moon's surface, contributing to our broader knowledge of lunar geology.

\section*{CRediT authorship contribution statement}
F. Thoresen performed conceptualization, methodology, software, validation, formal analysis, investigation, data curation, writing - original draft, writing - review \& editing, visualization, and project administration. 
I. Drozdovskiy performed conceptualization and writing - review \& editing. 
A. Cowley performed conceptualization, supervision and writing - review \& editing. 
M. Laban performed writing - original draft and writing - review \& editing.
S. Besse performed data curation and writing - review \& editing.
S. Blunier performed writing - review \& editing.

\section*{Acknowledgements}
We would like to express our sincere gratitude to the European Space Agency for their funding, which made this project possible. Special thanks to ExPeRT \& Spaceship EAC for their invaluable guidance throughout the course of the research. We are also grateful to the Human and Robotic Exploration group for their support and insightful guidance, with particular acknowledgement to Thomas Sheasby for his support. Acknowledgements are extended to the M3 team for their contribution in providing the essential data through the PDS3 node.

\bibliographystyle{apsrev4-1}
\bibliography{bib}

\end{document}